\documentclass[preprint,onecolumn,showpacs,preprintnumbers,amsmath,amssymb]{revtex4}
\usepackage{graphicx}
\usepackage{slashbox}

\begin{document}

\title{Entanglement of two interacting bosons in a two-dimensional isotropic harmonic trap}

\author{B. Sun and M. Pindzola}

\affiliation{Department of Physics, Auburn University, Auburn,
Alabama 36849, USA}

\begin{abstract}
We compute the pair entanglement between two interacting bosons in
a two dimensional (2D) isotropic harmonic trap. The interaction
potential is modelled by a 2D regularized pseudopotential. By
analytically decomposing the wave function into the single
particle basis, we show the dependency of the pair entanglement on
the scattering length. Our results turn out to be in good
agreements with earlier results using a quasi-2D geometry.
\end{abstract}


\maketitle

The rapid development of laser cooling, trapping and manipulating
ultracold bosonic atoms in optical lattices has attracted much
interest due to the capability of a precise control over the
number of atoms in each lattice site
\cite{Greiner,Meschede,Rempe,bloch,weiss}, thus raises hope for
the application of optical lattice systems in quantum information
and communication. Such lattice bosons exhibit novel properties
such as superfluid-Mott insulator quantum phase transition, which
has been observed successfully in the laboratory
\cite{Greiner,Meschede,Rempe}. On the other hand, recently there
has been growing interest in the interacting atomic system in low
dimensions \cite{bloch,weiss,esslinger,rad,citro,zoller,lozo}. In
ultracold atomic gases, low dimensions are usually obtained by
imposing strong confinement in some direction whose motional
degree of freedom is frozen, thus only virtual transitions are
allowed in that direction. For example, in a pioneer work of M.
Olshanii, it has been shown that the effective one dimensional
(1D) scattering length can be renormalized from a three
dimensional (3D) scattering length \cite{olshanii1}, opening
intensive discussions on the subject of restricted scattering. One
unique property possessed by 1D hard core bosonic systems is the
so called ``Bose-Fermi duality" predicted long time ago
\cite{girardeau}, which is recently verified in ultracold atomic
gases experimentally \cite{bloch,weiss}.

Quantum entanglement is a central concept in quantum mechanics. It
is viewed as valuable resources enabling the unmatched power of
quantum computation and communication \cite{Ni}. In previous
studies, the entanglement of $s$-wave scattering between two
interacting bosons has been investigated in a 1D harmonic trap
\cite{bo} and a 3D spherical harmonic trap \cite{law}. However,
the direct calculation from the 2D pseudopotential is still
absent. The recent work of using a 3D cylindrical harmonic trap
has revealed the entanglement properties under the quasi-2D
geometry \cite{bof}. It is natural to speculate that, for very
strong confinement, results from quasi-2D geometry calculations
should give essentially the same answer as the true 2D case. It is
the purpose of this paper to discuss the true 2D case and compare
it with quasi-2D results.

In this paper, we study the entanglement between two interacting
bosons in a 2D isotropic harmonic trap. We adopt the 2D
pseudopotential to model the $s$-wave scattering interaction. From
the obtained spectrum, we then compute the pair entanglement for
the ground state and the first two excited states perturbed by the
interaction. Our results turn out to be in good agreements with
earlier results using quasi-2D geometry, thus confirming the above
speculation.

The model we consider consists of two interacting bosons in a 2D
isotropic harmonic trap. The total Hamiltonian is
\begin{eqnarray}
H=\sum_{i=1}^{2}\left(-{\hbar^2 \over 2m}\nabla^2_{\vec
\rho_i}+{1\over 2}m\omega^2 \rho_i^2\right)+ V_{\rm s }({\vec
\rho}_1-{\vec \rho}_2),\label{bh}
\end{eqnarray}
where ${\vec \rho}_i=(x_i,y_i)$ is the position vector of the
$i$-th atom in cartesian coordinate. $V_{\rm s }$ is the $s$-wave
scattering potential given by \cite{olshanii}
\begin{eqnarray}
V_{\rm s }(\vec \rho_1-\vec \rho_2) = -{2\pi \hbar^2 \over m}
{1\over {\rm ln}(\Lambda a)}\delta^{(2)}(\vec \rho_1-\vec \rho_2)
\nonumber\\
\times \left(1-{\rm ln}(\Lambda \rho_{12})\rho_{12} {\partial
\over \partial \rho_{12}}\right),
\end{eqnarray}
with $a$ being the $s$-wave scattering length and $\rho_{12}\equiv
|\vec \rho_1-\vec \rho_2|$. $\Lambda$ is an arbitrary constant,
which will be cancelled out if the potential acts on the proper
wave functions with logarithm divergence($\propto {\rm
ln}(\rho)$). Similar to the 3D case, the regularization in the
above pseudopotential is necessary to avoid the unwanted
divergence.

Following the usual procedure of separating the total Hamiltonian
into center of mass (c.m.) and relative (rel) motion, we write
$H=H_{\rm c.m.}+H_{\rm rel}$, with $H_{\rm c.m.}$ being a 2D
harmonic oscillator. We further use $a_{\rm ho}=\sqrt{\hbar /(m
\omega})$ and $\hbar \omega$ as the length and energy scale,
respectively. Thus, the corresponding coordinate transformations
are ${\vec \varrho} =({\vec \rho}_1+{\vec \rho}_2)/\sqrt{2}$ and
${\vec \rho} =({\vec \rho}_1-{\vec \rho}_2)/\sqrt{2}$, for the
c.m. and rel-motion respectively. The dimensionless version of the
relative Hamiltonian is
\begin{eqnarray}
H_{\rm rel} = -{1\over 2}\nabla^2_{\vec \rho}+{1\over 2}\rho^2 +
V_{\rm s}(\vec \rho),
\end{eqnarray}
where $V_{\rm s}$ is now given by
\begin{eqnarray}
V_{\rm s}(\vec \rho) = -{\pi \over {\rm ln}(\Lambda
a)}\delta^{(2)}(\vec \rho)\left(1-{\rm ln}(\sqrt{2}\Lambda
\rho)\rho {\partial \over
\partial \rho}\right).
\end{eqnarray}
For energy $E=k^2$, the Wigner threshold law requires $ka\ll 1$ in
the free space. Applying to our case, this condition turns out to
be $a\ll a_{\rm ho}$ if the threshold law still holds in an
external trap. Otherwise, we must use an energy-dependent
scattering length to obtain more accurate results \cite{juli}.
Here we simply treat the scattering length as an independent
parameter. Although such a simplification is supposed to be valid
only for a small scattering length, the direct extension to a
large scattering length beyond the Wigner threshold law allows for
a closed form solution and a qualitative analysis for the current
problem. While for an energy dependent scattering length, it can
be only be obtained numerically through a self consistent
procedure and therefore lacks an analytical treatment.

From the relative Hamiltonian, we can see that the harmonic
orbitals with angular quantum number $m\neq 0$ are obvious
solutions because of the contact form of the scattering potential.
To obtain the non-trivial solutions with scattering length
dependence, we expand the eigenfunction into harmonic orbitals
with vanishing angular quantum number as in Ref. \cite{wilken},
$\psi(\vec \rho)=\sum_{n}\alpha_{n}\phi_{n}(\vec \rho)$, with the
2D $s$-wave harmonic orbital given by
\begin{eqnarray}
\phi_n(\vec \rho) = {1\over \sqrt{\pi}} e^{-\rho^2/2}L_n(\rho^2).
\end{eqnarray}
Here $L_n(x)$ is the $n$-th order Laguerre polynomial. The
expansion coefficient takes a simple form as $\alpha_{n}\propto
1/(2n+1-E)$, from which we can determine the corresponding
spectrum. After some algebra, we obtain the eigenfunction
\begin{eqnarray}
\psi_{\nu}(\vec \rho) = { \Gamma(-\nu) \over
\sqrt{\pi\zeta(2,-\nu)} } e^{-{\rho^2\over 2}} U(-\nu,1,\rho^2)
\end{eqnarray}
with eigenenergy $E=2\nu+1$. Here $\Gamma(n)$ is the complete
Gamma function and $U(a,b,x)$ is the confluent hypergeometric
function of the second kind. $\zeta(s,z)=\sum_{n=0}^\infty
1/(n+z)^s$ is the Hurwitz zeta function \cite{arfken}.

For $\rho\to 0$, we have the asymptotic results as follows
\cite{note}
\begin{eqnarray}
&& U(-\nu,1,\rho^2) \to B_{\nu}{\rm ln}\rho + C_{\nu}, \nonumber\\
&& B_{\nu} = -{2\over \Gamma(-\nu)}, \nonumber\\
&& C_{\nu} = { 1+ 2\gamma\nu+{\mathcal F}(1-\nu)\nu \over
\Gamma(1-\nu)},\label{asmpU}
\end{eqnarray}
where ${\mathcal F}$ is the digamma function \cite{arfken}. Thus
the wave functions are consistent with the scattering potential
and indeed give a result independent of $\Lambda$.

\begin{figure}[htb]
\centering
\includegraphics[width=3.25in]{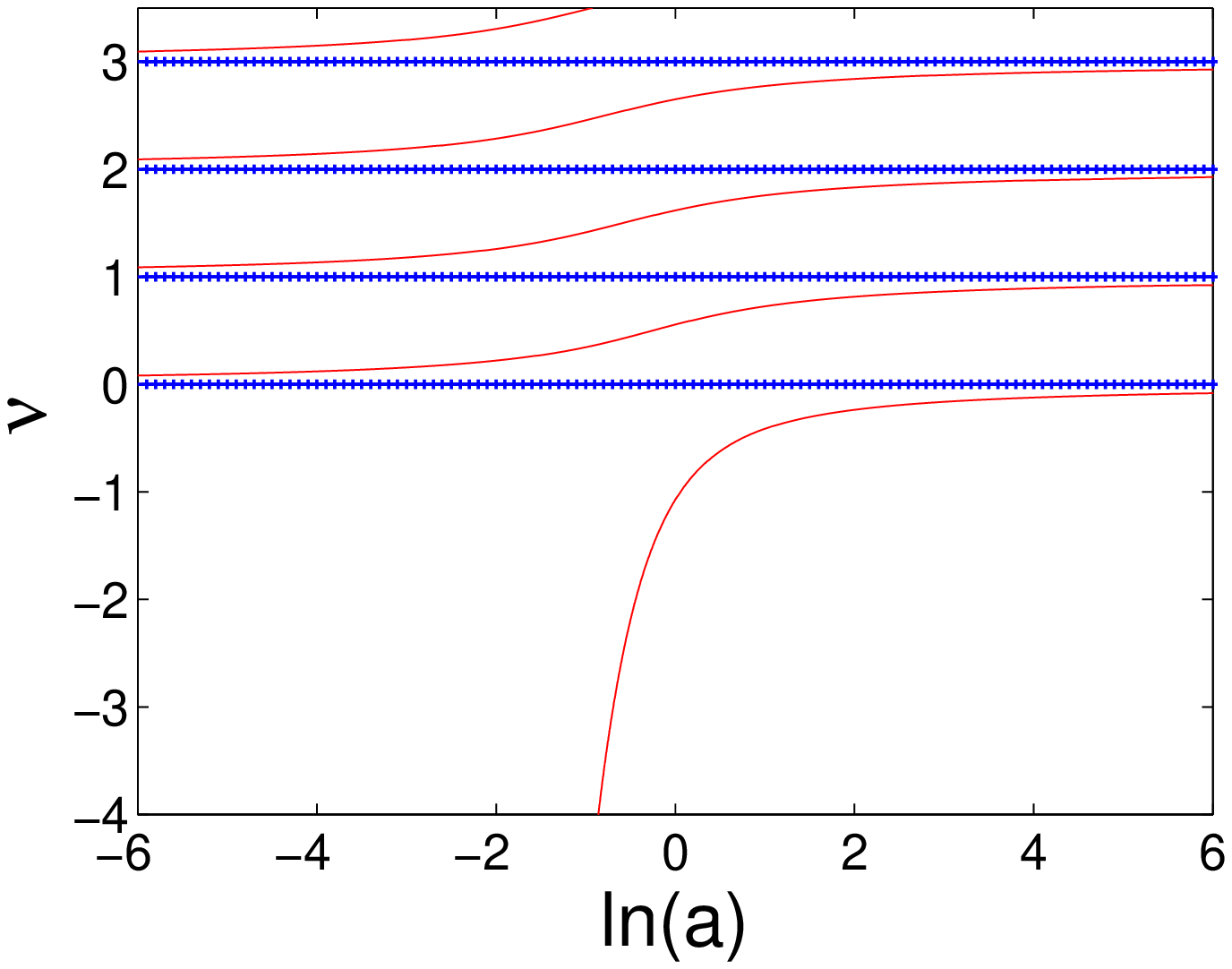}
\caption{Energy spectrum $\nu$ as function of ${\rm ln}(a)$. The
unmarked red curves are for those solutions affected by the
scattering potential, and the marked blue horizontal lines are for
asymptotic values as ${\rm ln}(a)\to \pm\infty$.} \label{va2D}
\end{figure}

The energy quantization condition is found to be
\begin{eqnarray}
a &=& \sqrt{2} {\rm exp}\left[{C_{\nu}\Gamma(-\nu)\over 2}\right]\nonumber\\
&=& \sqrt{2} {\rm exp}\left[-\gamma-{{\mathcal F}(-\nu)\over
2}\right] . \label{eq}
\end{eqnarray}
We note the scattering length is nonnegative \cite{verhaar}. The
results of $\nu$ as function of ${\rm ln}(a)$ are shown in Fig.
\ref{va2D}. We find that the energy spectrum behaves like that in
a 1D and 3D trap: (1) the energy is a smooth and monotonic
function of the scattering length, and (2) the energy is bounded
except for the ground state.

\begin{figure}[htb]
\centering
\includegraphics[width=3.25in]{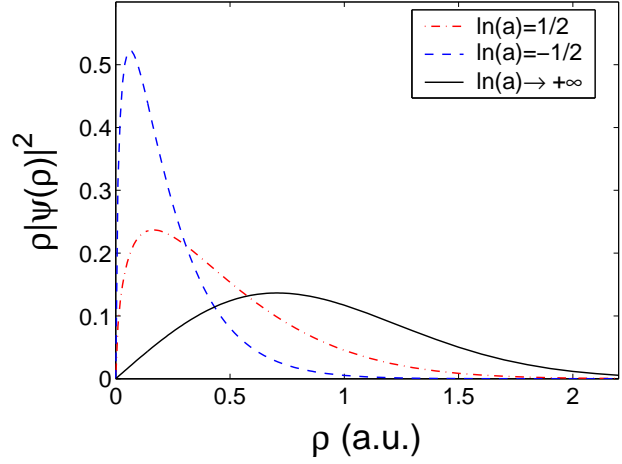}
\caption{The radial density distribution of the ground state.}
\label{WF}
\end{figure}

The radial densities $\rho|\psi(\rho)|^2$ of the ground state is
shown in Fig. \ref{WF} for chosen scattering lengths ${\rm
ln}(a)=\pm 1/2$ and ${\rm ln}(a)\to +\infty$. We can see that the
atoms tend to approach each other as ${\rm ln}(a)$ decreases. In
addition, we note that the derivative diverges at $\rho=0$ for
${\rm ln}(a)=\pm 1/2$ which originates from the scattering
potential. While for ${\rm ln}(a)\to +\infty$ it has a finite
derivative corresponding to the non-interacting case. This is
because in this limit, the effective coupling strength $g_{\rm
2D}\simeq -2\pi/{\rm ln}(a)$ approaches 0 \cite{calarco}.

After obtaining the spectrum, we are able to compute the
entanglement. We adopt von Neumann entropy as our pure state
entanglement measure \cite{Ni}. We are interested in the pair
entanglement for pure states with c.m. motion being the ground
state of the trap. The total wave function is
\begin{eqnarray}
\Psi_{0\nu}(\varrho,\rho) = { \Gamma(-\nu)\over \pi
\sqrt{\zeta(2,-\nu)} } e^{-{\varrho^2+\rho^2\over
2}}U(-\nu,1,\rho^2).
\end{eqnarray}
First, we briefly review the method in Ref. \cite{law} which is
also applicable to our case. Since
$\varrho^2=(\rho_1^2+\rho_2^2+2\rho_1\rho_2 {\rm
cos}(\varphi_1-\varphi_2))/2$ and
$\rho^2=(\rho_1^2+\rho_2^2-2\rho_1\rho_2 {\rm
cos}(\varphi_1-\varphi_2))/2$, we can expand the total wave
function as $\Psi_{0\nu}(\varrho,\rho)=\sum_m A_m(\rho_1,\rho_2)
e^{im(\varphi_1-\varphi_2)}/(\rho_1 \rho_2)$. $A_m(\rho_1,\rho_2)$
can be further written in the Schmidt form
$A_m(\rho_1,\rho_2)=\sum_s
\kappa_{m,s}\chi_s^{(m)}(\rho_1)\chi_s^{(m)}(\rho_2)$ with
\begin{eqnarray}
\int_0^{+\infty}A_m(\rho_1,\rho_2)\chi_s^{(m)}(\rho_2)d\rho_2=\kappa_{m,s}\chi_s^{(m)}(\rho_1).
\end{eqnarray}
Thus, the total wave function can also be written in the Schmidt
form
\begin{eqnarray}
\Psi_{0\nu}(\varrho,\rho) = \sum_{m,s}\kappa_{m,s} \left[
{\chi_s^{(m)}(\rho_1) \over \rho_1} e^{im\varphi_1}\right]
\left[{\chi_s^{(m)}(\rho_2) \over \rho_2}
e^{im\varphi_2}\right]^*.
\end{eqnarray}
The entropic entanglement can then be calculated from the Schmidt
coefficients as $\mathcal{E}=-\sum_{m,s} \kappa_{m,s}^2{\rm
ln}(\kappa_{m,s}^2)$.

\begin{figure}[htb]
\centering
\includegraphics[width=3.25in]{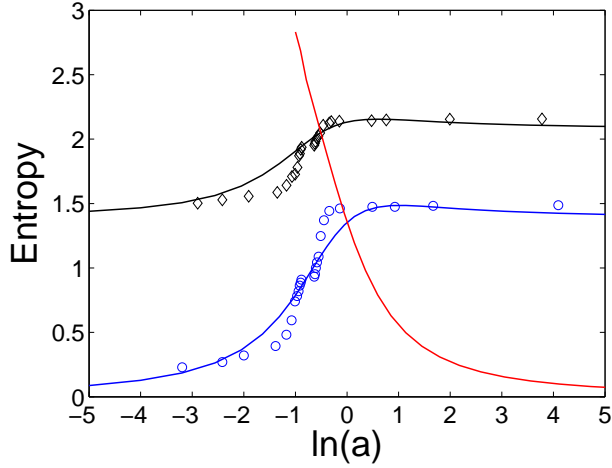}
\caption{Von Neumann entropy as function of the scattering length.
The (red) decreasing curve is for the ground state. The lower
(blue) and upper (black) of the two increasing curves are for the
first and second excited state (both perturbed by interaction) in
the relative motion. The c.m. motion is always assumed in the
ground state of the trap. Data points with circles (blue) and
diamonds (black) are from quasi-2D calculations with $\eta=20$ and
an effective scattering length of Eq. (\ref{aeff}). See text for
details.} \label{ent2Dt}
\end{figure}

Here we use an alternative method to calculate the entanglement.
We find the following decomposition in the single particle basis
\begin{eqnarray}
&&\Psi_{0\nu}(\vec \rho_1,\vec \rho_2) = \sum_{\vec m,\vec n}
C_{\vec m \vec
n}\phi_{\vec m}(\vec\rho_1)\phi_{\vec n}(\vec \rho_2)\nonumber\\
&=&{2\over \pi \sqrt{\zeta(2,-\nu)} } \sum_{\vec
m,\vec n}{1\over 2\nu-\sum_{j}(m_j+n_j)} \prod_{j} {1+(-1)^{m_j+n_j}\over 2}\nonumber\\
&& \times (-1)^{m_j-n_j\over 2} { \Gamma\left({m_j+n_j+1\over 2}
\right) \over \sqrt{m_j!n_j!} }
\phi_{m_j}(\rho_{1j})\phi_{n_j}(\rho_{2j}),
\end{eqnarray}
where $j=\{x,y\}$, ${\vec m}=(m_x,m_y)$ and ${\vec n}=(n_x,n_y)$.
$\rho_{kj}$ is the $j$-th component of the position vector $\vec
\rho_{k}$ for the $k$-th atom in cartesian coordinate, {\it e.g.},
$\rho_{1x}=x_1$. $\phi_{\vec m}(\vec \rho)$ is the 2D harmonic
orbital again in cartesian coordinate. This decomposition can be
proved straightforwardly and its validity holds as long as the
c.m. motion is in the ground state of the trap. We then
diagonalize the expansion coefficient matrix $[C_{\vec m \vec n}]$
and obtain $U^{\dagger}CU=D$, where $D$ is a diagonal matrix with
diagonal elements $\lambda_n$ satisfying $\sum_n \lambda_n^2=1$.
The von Neumann entropy is then given by $\mathcal{E}=-\sum_n
\lambda_n^2{\rm ln}(\lambda_n^2) $. The calculated entropy is
shown in Fig. \ref{ent2Dt} where the (red) decreasing curve is for
the ground state and the lower (blue) and upper (blue) increasing
curves are for the first and second excited state, respectively.
We find that the tendency of the curves remains essentially the
same as in the 3D case \cite{law}. In addition, the result is in a
qualitative agreement with that from a cylindrical trap with an
aspect ratio $\eta=\omega_{\perp}\ll \omega_z$ \cite{bof}. It is
interesting to make a quantitative comparison between a 2D and
quasi-2D geometry. To do this, we note that, in the quasi-2D case,
an effective scattering length is given by \cite{olshanii,shl}
\begin{eqnarray}
a_{\rm eff}=2.092 a_z {\rm exp}\left(-\sqrt{\pi\over 2} {a_z\over
a_{3D}} \right), \label{aeff}
\end{eqnarray}
where $a_{3D}$ is the original 3D scattering length and
$a_z=\sqrt{\hbar/(m\omega_z)}$ is the length scale along the
tightly confined $z$-axis. Transforming Eq. (\ref{aeff}) into the
unit that is adopted in the current paper, it reads
\begin{eqnarray}
{\rm ln}{a_{\rm eff}\over a_{\perp}} = {\rm ln}{2.092 \over
\sqrt{\eta} } -\sqrt{\pi\over 2} {a_z\over a_{3D}}. \label{a2}
\end{eqnarray}
Our comparison with $\eta=20$ is also shown in Fig. \ref{ent2Dt}
where data points with diamonds (circles) are for the first
(second) excited state. We do not make comparison for the ground
state \cite{nr}. We can see that the two results agree well with
each other except for the critical region where $a_{3D}\sim \pm
\infty$ and the corresponding effective scattering length is ${\rm
ln}a_{\rm eff}^{(cr)}={\rm ln}(2.092/\sqrt{\eta})$ as from Eq.
(\ref{a2}). For $\eta=20$, ${\rm ln}a_{\rm eff}^{(cr)}\simeq
-0.76$. The 2D results vary more slowly than the quasi-2D results
around the critical region. The discrepancy is expected since in
deriving the effective scattering length in Eq. (\ref{aeff}),
there is no trapping potential in the transverse direction. While
in our study, the existence of such a transverse trapping
potential may have an effect on the scattering problem.

\begin{figure}[htb]
\centering
\includegraphics[width=3.25in]{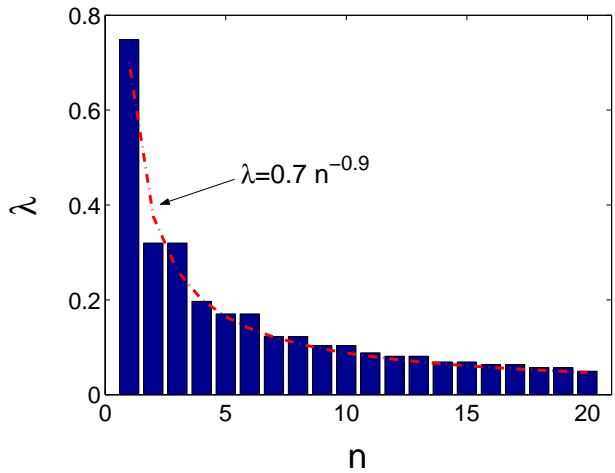}
\caption{The first 20 eigenvalues $\lambda_n$ in descending order
for the ground state with ${\rm ln}(a)=-0.5359$. The (red) dashed
dotted line is an exponential fit.} \label{lambda}
\end{figure}
\begin{figure}[htb]
\centering
\includegraphics[width=3.25in]{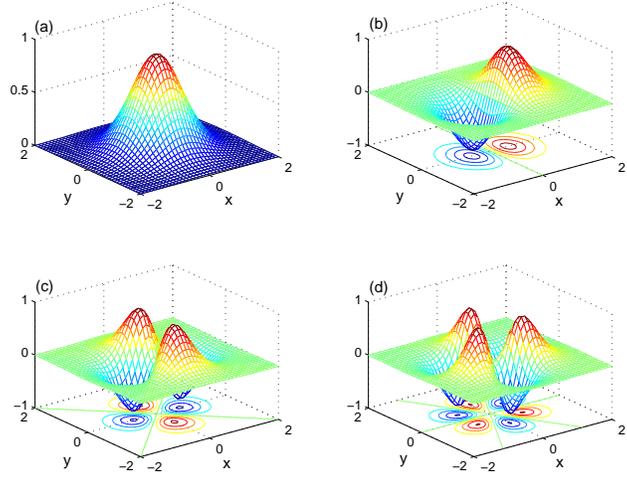}
\caption{Several Schmidt orbitals for the ground state with ${\rm
ln}(a)=-0.5359$. From (a)$\to$(d), the corresponding eigenvalues
are 0.7408,0.3164,0.1686, and 0.1029, respectively.}\label{obt}
\end{figure}

To better understand the entanglement properties, we give an
example for the eigenvalues and eigenfunctions. We randomly choose
a scattering length ${\rm ln}(a)=-0.5359$ and the corresponding
results are shown in Figs. \ref{lambda} and \ref{obt},
respectively.

In Fig. \ref{lambda}, we show the first 20 eigenvalues for the
ground state. Using a fitting curve $\lambda_n=\alpha n^\beta$, we
obtain $\alpha\simeq 0.7$ and $\beta\simeq -0.9$. This slow
decrease is expected since the eigenfunction $\Psi_{0\nu}(\vec
\rho_1,\vec \rho_2)$ is singular at the origin which can only be
removed by a regularized pseudopotential. On the other hand, the
entanglement is convergent which can be seen from the power law
relation of the eigenvalues and further confirmed by our numerical
results. In Fig. \ref{obt}, we show the eigenfunctions (Schmidt
orbitals) for four different eigenvalues for the ground state. The
Schmidt orbital $\tilde{\phi}(\vec\rho)$ is related to the
harmonic basis by a unitary transformation, {\it i.e.},
$\tilde{\phi}(\vec\rho)\equiv U^{\dagger}\phi(\vec \rho)$. The
${\vec p}$-th component of $\tilde{\phi}(\vec\rho)$ can be written
as $\tilde{\phi}_{\vec p}(\vec\rho)=\sum_{\vec m}U^{\dagger}_{\vec
p \vec m}\phi_{\vec m}(\vec \rho)$. Compared to harmonic orbitals,
these orbitals experience deformation due to the interaction. On
the other hand, the symmetry of the orbitals is still preserved
due to the isotropic interaction. For example, if we rotate the
orbital in Fig. \ref{obt}(b) by $90^o$, we obtain another
eigenstate which is degenerate to it. These two eigenstates
correspond to the second and third eigenvalue in Fig.
\ref{lambda}.

%

Two limiting cases, {\it i.e.}, ${\rm ln}(a)\to \pm\infty$,
deserve a simple discussion. When ${\rm ln}(a)\to +\infty$, we
have already mentioned before that it corresponds to the
non-interacting regime. While for ${\rm ln}(a)\to -\infty$, we can
see that the entanglement goes without limit. We identify it as
the bound state regime, where the binding energy is calculated
from Eq. (\ref{eq}) with the result $E_b\to -2\nu \simeq 1/a^2$ as
$\nu\to -\infty$. The size of the bound state is on the order of
$a$, thus a vanishingly small cross section in this regime. This
is a quite general conclusion in the scattering theory. At low
energy, the scattering amplitude for an $l$-wave is given by
\cite{llbook}
\begin{eqnarray}
f_l \sim {(kb)^{2l} \over -a_l^{-1}-r_l k^2 /2 - i(kb)^{2l} k},
\end{eqnarray}
where $b$ is a constant only depending on the scattering potential
and $r_l$ is the effective range. The bound state can be obtained
from the pole of $f_l$. For low energy ($k\to 0$) s-wave
scattering ($l=0$), it gives $k=i/a$. The corresponding wave
length $\lambda\sim |1/k| =a$, thus a size of $a$. The bound state
energy is $E_b=|k^2|=1/a^2$. Both analysis are in agreement with
our results.

In conclusion, we have carried out the calculation for the
entanglement between two interacting bosons in a 2D isotropic
harmonic trap. The $s$-wave scattering is modelled by the
well-known 2D regularized pseudopotential. By decomposing the wave
function into single particle basis, we have shown the dependency
of the pair entanglement on the scattering length. In addition, we
have analyzed the Schmidt mode (natural orbit) of the system. Our
current study only focuses on pure states, the more general
entanglement properties with a temperature dependence will be an
open question for future investigations \cite{notes1}. With the
possible realization of a 2D few-body quantum system in the future
and the controllability of the scattering length through Feshbach
resonance, we hope our results could be helpful to the study of
quantum information and communication in low dimensional systems.

This work is supported by grants with NSF.

\newpage

\begin{thebibliography}{99}
\bibitem{note}
For $\rho\to 0$, we can use $dU(-\nu,1,\rho^2)/d\rho^2=\nu
U(-\nu+1,2,\rho^2)\to \nu [1/(\Gamma(1-\nu)\rho^2)+O(1)]$ and
integrate on both sides to obtain $U(-\nu,1,\rho^2)\to B_{\nu}{\rm
ln}(\rho)+C_{\nu}$. Then $C_{\nu}$ can be obtained by using the
Wronskian of two confluent hypergeometric functions
$M(-\nu,1,\rho^2)dU(-\nu,1,\rho^2)/d\rho^2-UdM/d\rho^2\to
-(1+1/\rho^2)/\Gamma(-\nu)$. Comparing both sides, we immediately
obtain Eq. (\ref{asmpU}).
\bibitem{nr}
The ground state entanglement in a quasi-2D geometry is
computationly intensive. Even for a moderate scattering length, it
requires much more basis for the expansion than the 2D case in
order to achieve convergence. Due to this numerical difficulty,
the ground state entanglement is not given in Ref. \cite{bof} and
as a result, a comparison is not carried out.
\bibitem{notes1}
For the thermal entanglement which is related to mixed states in
general, we can use negativity as entanglement measure
\cite{werner}. Different from pure state calculation which only
need the reduced density matrix, the calculation of negativity
need the full two-body density matrix. This is a computationally
demanding task due to the very large Hilbert space.
\end{thebibliography}
\end{document}